\documentclass[12pt,preprint]{aastex}
\usepackage{amsmath}
\usepackage{amssymb}
\usepackage{latexsym}
\usepackage{psfig}
\usepackage{graphicx}
\usepackage{color}
\usepackage{ulem}

\shorttitle{Detecting Inspiraling Binary Systems}
\shortauthors{authors}
\begin{document}

\title{Gravitational Waves from Extragalactic Inspiraling Binaries:
  Selection Effects and Expected Detection Rates} 

\author{Philip Nutzman\altaffilmark{1}, Vicky
  Kalogera\altaffilmark{1}, Lee Samuel Finn\altaffilmark{2}, \\ Cy
  Hendrickson\altaffilmark{1}, Krzysztof Belczynski\altaffilmark{1,3}}

\affil{$^{1}$Northwestern University, Dept. of Physics \& Astronomy,
  2145 Sheridan Rd., Evanston, IL 60208\\ $^{2}$Center for
  Gravitational Wave Physics, The Pennsylvania State University,
  University Park, PA 16802\\ $^{3}$Lindheimer Postdoctoral Fellow\\
  me, vicky, c-hendrickson, belczynski@northwestern.edu; and
  LSFinn@PSU.Edu}

\begin{abstract} 

We examine the selection effects that determine how the 
population of inspiraling binary compact objects (BCOs) is reflected by those potentially observed with ground-based interferometers
like LIGO.  We lay the ground-work for the interpretation of future 
observations in terms of constraints on the real population and,
correspondingly, binary star evolution models.

To determine the extra-galactic population of inspiraling binaries
we combine data on distance and blue luminosity from galaxy catalogs with current models of the galactic BCO mass distribution to simulate the physical distribution of binaries in the nearby universe. We use Monte Carlo methods to determine the
fraction of binaries observable by the LIGO
detectors from each galaxy as a function of the BCO chirp mass.
We examine separately the role of source distance, sky position,
time of detection, and binary system chirp mass on detection
efficiency and selection effects relevant to the three LIGO
detectors.  Finally, we discuss the implications of the nearby
geography of space on anticipated GW detection and compare our
results to previous studies, which have assumed uniform galaxy
volume density and fixed chirp mass for binary compact objects.

From these considerations, actual BCO inspiral observations or significant upper limits on the coalescence rate anticipated in the near future by ground-based interferometers can be used to improve our
knowledge of the galactic binary inspiral rate and to constrain models
of radio pulsar characteristics and binary star evolution channels
leading to neutron star or black hole binaries.

\end{abstract}


\section{Introduction}\label{sec:intro}

Binary compact objects (BCO) with neutron stars or black holes hold a special place among gravitational wave (GW) sources. The discovery of
PSR~B1913+16~\citep{ht}, the first binary pulsar system, inspired the
detailed study of inspiraling compact binaries and provided the first
observational evidence for the existence of gravitational
radiation~\citep{TW89}. Binary systems like PSR~B1913+16 are driven to
coalescence by a GW emission catastrophe: in the last approximately
20\,s before they coalesce they radiate their remaining binding energy
(approximately $2\times10^{52}$~ergs) as gravitational waves in a band
accessible to the large ground-based detectors like the Laser
Interferometer Gravitational Wave Observatory~\citep[LIGO;][]{ligoR}
and VIRGO~\citep{virgo}\footnote{Somewhat less energy will be radiated
  in the bands accessible GEO \citep{ligoR} and TAMA \citep{TAMA}, and
  less still in the bands accessible to the bar detectors AURIGA
  \citep{AURIGA}, ALLEGRO \citep{ALLEGRO}, EXPLORER
  \citep{EXPLORER1,EXPLORER2} and NAUTILUS \citep{NAUTILUS}.}.
  
Current observational constraints on the population of neutron star or
black hole binary systems depend on radio pulsar observations of just
a handful of Galactic binary systems \citep[e.g.,][]{burgay,kal03}. In
contrast, the LIGO and VIRGO detectors will observe stellar mass
inspiraling BCOs at extra-galactic distances. They will also be sensitive to black hole binaries, which
are not observable electromagnetically. Correspondingly, observations
by this new generation of detectors can help constrain the binary
coalescence rate density in the nearby Universe, and binary evolution
models for the formation of such sources, in ways not possible with
electromagnetic observations alone. In this work we begin laying the
ground-work for the astrophysical interpretation of future GW
observations of BCO inspiral by focusing attention on
the selection effects associated with GW observations: in particular, effects associated
with binary component masses, the GW antenna beam and
the local geography of the Universe.

Over the past decade there have been many predictions of the
detection rate in LIGO of BCO inspirals \citep[e.g., ][]{belc02,kal03}.  These calculations all begin by
estimating the Galactic coalescence rates and extrapolating it to 
other galaxies. The observed rate is then calculated assuming that
galaxies (and, thus, binaries) are distributed homogeneously and
isotropically in the local universe, and that the LIGO detector
observes all coalescing binaries inside a fixed distance, which is the
radius of a sphere that would have the same volume as is effectively
surveyed by the detector. These approximations are inadequate when we
wish to go beyond order-of-magnitude predictions and actually
interpret observed events as constraints on the actual compact object
binary population, as is our goal. 

To improve on past models for the physical population of inspiraling
compact binary systems, we use galaxy catalogs to model the actual
distribution of galaxies in the local universe and we use stellar synthesis calculations \cite[specifically those of ][]{belc02} to model the mass distribution of binaries within each galaxy. From the constructed population models, we determine the compact binary coalescence rate and distribution with binary system mass that we expect the LIGO detector system to observe, taking full account of each galaxy's distance and declination, the LIGO detector system's noise spectrum, and its position and orientation on Earth. 

Our principal goal in relating a physical population model to the
distribution that we expect modern GW detectors to observe is to
enable observations by those detectors to constrain the population
model (see \citealt{bulik} and \citealt{bulik2} for recent studies with similar goals). Through the calculations described here, comparison of future observed rates or rate upper limits constrain stellar synthesis models and the overall binary compact object population.  While our principal
interest is in preparing for this kind of interpretation of
forthcoming observations, as a by-product of our investigations we
have improved detection rate predictions as well. 

In \S\ref{sec:background} we present an overview of the various
approaches used so far for the extrapolation of Galactic detection
rates to extragalactic distances and introduce our novel galaxy-by-galaxy approach, whereby we calculate the detectability of BCO inspiral for each galaxy in our catalog.  In \S\ref{sec:method} we
describe how we calculate, from the detailed extra-galactic population
model described in \S\ref{sec:background}, the observed distribution
of BCOs.  In \S\ref{sec:results}
we discuss our results, including the LIGO detector system's
efficiency for detecting binaries from different galaxies in the
nearby universe, the expected observed coalescing binary mass
distribution, new detection rate predictions, and the implications of
the geography of the nearby univese for detection of binary compact
object systems. We end in \S\ref{sec:conclusions} with a summary of
our main conclusions.

\section{Estimating the Binary Compact Object Population Density in the Nearby Universe}
\label{sec:background}

Our final goal is to understand how the actual population and
distribution of coalescing binary compact objects (BCOs) is mapped to 
GW observations by LIGO (or GEO, TAMA or VIRGO) of
inspiraling binary systems. The signal amplitude
 depends on the system's distance, location on the detector's sky,
orbital plane inclination with respect to the detector line-of-sight,
and the binary chirp mass (cf.\ eq.[\ref{eq:chirpMass}]). In this
section we characterize the compact object binary population
distribution in space and component masses: i.e., the density as a
function of distance, position on the detector sky, and the chirp
mass.

\subsection{Previous Studies}\label{sec:previous}

All studies of compact object coalescence GW detection rates begin
with an estimate of the intrinsic coalescence in our own Galaxy.  Two
methods have been employed to derive these rates: statistical analysis
of the observed sample of inspiral binary candidates and theoretical
investigations based on understanding of BCO
formation. The first approach so far provides us with the best
rate constraints (see \cite{burgay} and \cite{kal03} for the most
recent estimates in view of a newly discovered NS-NS system), but is
still limited by a small observed sample (currently 3 systems), some
uncertainty in the selection effects associated with, e.g., pulsar
beaming and radio luminosity, and by the absence of known BH-NS and
BH-BH binaries.  The second approach provides us with results for all
types of BCOs based on binary population synthesis models, which are
generally calibrated to match the empirical estimates of Type II
supernova rates \citep[hereafter BKB and references therein]{belc02}, but the uncertainties associated with binary evolution models are significant. At present our best estimates
of Galactic coalescence rates place them in the range of
$10^{-5}-10^{-3}$\,yr$^{-1}$ for NS-NS, $10^{-6}-10^{-4}$\,yr$^{-1}$
for BH-NS, and $0-10^{-4}$\,yr$^{-1}$ for BH-BH.

All extrapolations from Galactic rates to extragalactic rates are based
on the assumption that the formation of BCOs in a
region is proportional to the blue luminosity in that region,
corrected for reddening \citep{phinney}. Correspondingly, the
coalescence rate density $R_{\mathrm{det}}$ within a distance 
$D_{\max}$ is proportional to the Milky Way coalescence rate dentsity
$R_{\mathrm{MW}}$ and the ratio of the blue luminosity within  the
sphere to the Milky Way blue luminosity:
\begin{equation}\label{eq:old}
 R_{det} = (\frac{\mathcal{L}_{b}} {L_{\mathrm{MW}}})(\frac{4}{3}\pi
 D_{\max}^{3})\,R_{\mathrm{MW}}, 
\end{equation}
where $\mathcal{L}_{b}$ is the mean blue luminosity volume density
within a distance $D_{\max}$, $L_{\mathrm{MW}}$ is the total blue
luminosity of the Milky Way, and $R_{\mathrm{MW}}$ is the coalescence
rate in the Milky Way. The distance $D_{\max}$ is calculated so that,
if coalescing binaries with given component masses (e.g.,
1.4~$\mathrm{M}_\odot$) are assumed to be uniformly distributed
throughout space with rate density $\dot{n}$, the rate of detections
in a GW detector above a given signal-to-noise ratio
is 
\begin{equation}
\dot{N}_{\mathrm{GW}} = \frac{4\pi}{3}D_{\max}^3\dot{n};
\end{equation}
i.e., $D_{\max}$ is the radius of a sphere whose volume is the
effective volume surveyed by the detector for binary systems with
these component masses \citep{finn96,finn_proceed,kal01}. 

\subsection{A Galaxy Catalog Approach}\label{sec:approach}

Here we describe a method appropriate for extrapolating Galactic BCO
coalescence rates beyond the galaxy, for the purpose of estimating
inspiral detection rates in GW detectors. Our ultimate
goal is not the prediction of what might be seen, but to understand
how observed bounds on the coalescence rate, perhaps as a function of
the BCO component masses determined from these observations, can 
inform our understanding of binary populations in galaxies and binary
population synthesis.  Toward that end we improve on the method of
binary inspiral rate estimation, described in section
\ref{sec:previous}, in three ways:
\begin{enumerate}
\item The blue luminosity scaling argument is accurate when the survey
  volume is so large that local fluctuations in the galaxy density and
  the blue luminosity per galaxy are small. This is not the case in
  the local universe, where the galaxy distribution is strongly
  anisotropic. To overcome this difficulty we make use of galaxy
  catalogs to map the true distribution of galaxies in space and in
  blue luminosity.
\item The LIGO detectors sensitivity is not
  isotropic: e.g., it is more sensitive to binaries immediately above
  and below the plane of the detectors arms than to binaries in the
  detector plane. We take full account of the detector's actual
  beam and its orientation with respect to the sky, averaged over the
  sidereal day.
\item Estimates to date use a $D_{\max}$ chosen for a particular
  combination of component masses. \footnote{In previous surveys binaries are assumed to be
    either double neutron star, with component masses of
    $1.4~\mathrm{M}_\odot$, or black hole neutron star binaries, with
    component masses of $10~\mathrm{M}_\odot$ and
    $1.4~\mathrm{M}_\odot$.}  We estimate the inspiral detection rate
  for the theoretically expected distribution of component masses, as opposed to some canonical mass.
\end{enumerate}

\subsubsection{Spatial Distribution of Galaxies}
\label{sec:distribution}

Galaxies are distributed anisotropically in the local universe, and
this affects the interpretation of BCO inspiral detection rates in
terms of galactic populations of BCOs.  To take into account the
nearby distribution of galaxies, we draw galaxy data from two
catalogs: the Lyon-Meudon extragalactic database~\citep[hereafter
LEDA]{LEDA}, and the Tully Nearby Galaxy Catalog~\citep[hereafter NBG]{tully}.
\footnote{Previous work by \cite{Lipunov} similary utilizes the Tully Nearby Galaxy Catalog to map the distribution of gravitational wave sources.  Note that our purpose and analysis greatly diverges from this earlier work; our emphasis on selection effects demands different considerations (e.g., knowledge of detector location, orientation and noise characteristics, correct compensation for incompleteness of galaxies in Tully's NBG and galaxies beyond the range of NBG, etc) and enables us to consider different questions from those in Lipunov et al. }      
\normalcolor
We select these galaxy data sources for their homogeneous coverage
of the sky: GW detectors have some sensitivity to BCOs
in every direction on the sky and so data from surveys that cover only
sectors of the sky, such as the Sloan Digital Sky
Survey~\citep{SDSS}, are not adequate. 

Our principal source of galaxy data is the LEDA. It is comprehensive
for galaxies with blue magnitudes brighter than 14.5 and partially
complete for those as faint as $m=17$~\citep{LEDA}. Owing to the
difficulties in determining galaxy distances, LEDA lacks distance data
for all but a small fraction of its galaxies.  The NBG, on the other
hand, has excellent information for galaxies with radial velocity less
than 2500 km/s, but makes no attempt to be comprehensive beyond that
range~\citep{tully}. In order to have available the best galactic
distance estimates we use the NBG distance data with the corresponding
LEDA galaxies and derive distances for all other LEDA galaxies using
the measured radial velocities and $H_{0}$=
70\,km\,$s^{-1}$\,Mpc\,$^{-1}$.  Our synthesized catalog thus provides
the needed parameters of right ascension $\alpha$, declination
$\delta$, distance $D$, and blue luminosity corrected for reddening
for each galaxy.

A primary concern with any galaxy catalog is incomplete coverage of
faint galaxies.  Our catalog is complete for galaxies brighter than
blue magnitudes $14.5$; correspondingly, incompleteness is important
for faint galaxies beyond approximately 20 Mpc and for galaxies of
increasing intrinsic brightness at greater distances.

To compensate for the missing galaxies while still accounting for
their non-uniform spatial distribution we introduce a
distance-dependent correction factor, which is the ratio of the
expected blue luminosity, were the catalog is complete, to the blue
luminosity reported by the catalog at each distance.  To derive this
factor we calculate the expected distribution of luminosity with
distance assuming that the true distribution with distance is
proportional to the luminosity distribution at that distance for
galaxies {\it brighter} than the completeness at each distance.
Since our galaxy catalog is essentially complete with regards to
cumulative blue luminosity up to approximately 22~Mpc, we calibrate
the resulting distribution to match the catalogued distribution at
this distance.

Figure \ref{fig:lum_cum} shows the corrected cumulative blue luminosity as a function of distance. For comparison we show also the cumulative blue luminosity exected under the assumption of uniform blue luminosity density as adopted in~\cite{kal01}. The normalization of the uniform density blue luminosity curve is based on galaxy surveys out to large distances. Three elements of this figure are worth noting in more detail.  First, the uniform blue luminosity density approximation clearly underestimates the actual blue luminosity contributed by nearby galaxies. Second, the Virgo cluster contributes to a large step in the cumulative blue luminosity at approximatly 20~Mpc. The higher local blue luminosity density at this distance significantly increases the estimated detection rate as a fraction of the total number of coalescing systems, especially for initial detectors. Finally, beyond the Virgo cluster the blue luminosity grows more slowly than the third power of distance (see \S\,\ref{sec:cum}) as the local over-abundance of galaxies blends into the homogeneous distribution at larger scales.

An important assumption in our correction for completion is that the
spatial distribution of the missing luminosity follows the spatial
distribution of the recorded galaxies.  This approximation is best
where the fraction of uncatalogued luminosity is lowest: i.e., where
the missing luminosity is dominated by the recorded luminosity. At
large distances the opposite is true and we expect this approximation
to be less accurate.  Nevertheless, the specific nature of the spatial
distribution at these distances is negligible for initial LIGO, where
only small fractions of detections for initial LIGO will be due to
galaxies at these large distances.

\subsection{The chirp mass and its
  distribution}\label{sec:binary_prop}

The GW inspiral signal depends, to leading order, on the so-called chirp mass,
denoted $\mathcal{M}$: 
\begin{equation}
\mathcal{M} \equiv \mu^{3/5}M^{2/5},
\end{equation}
where $\mu$ is the system's reduced mass and $M$ its total mass. To
complete our specification of the physical compact object binary
population we must characterize the populations distribution with
chirp mass. 

We model the distribution of compact object binary systems with
component masses using the StarTrack (BKB) binary synthesis code. 
We use the resulting chirp mass distribution to determine the signal
strength from a given binary at our GW detector. 
For example, Figure \ref{fig:chirpdist} (which also appears in Bulik \& Belczynski 2003) 
shows the distribution of
BCO chirp masses as calculated using the StarTrack binary synthesis 
code and BKB reference model A. (While we confine attention to
reference model A in this section, we explore the dependence of the
observed rates and distributions on the full range of BKB reference
models in \S\ref{sec:cum}.)

\section{The distribution of BCOs anticipated in observations} 
\label{sec:method}

Here we use the model for the physical BCO population developed in the previous section to determine the observed coalescence rate and distribution as a function of the detector's noise characteristics.

The strength of the BCO inspiral signal observed in a detector system  is characterized by the
signal-to-noise ratio $\rho$ which depends on the detector's noise spectrum and the BCO chirp mass, distance, sky position, and orbital inclination. Detected binary systems will have
signal-to-noise greater than a threshold $\rho_0$. 

 For ground-based
GW detectors like LIGO the signal from a coalescing
compact object binary system persists in the detector for on-order
seconds. Over that period the accumulated signal-to-noise will depend
on the detector's noise spectrum and the binary system's distance,
orbital plane inclination with respect to the detector line-of-sight,
location on the detector's sky, and a function --- the so-called
``chirp mass'' (cf.\, eq.  \ref{eq:chirpMass}) --- of the binary's
component masses. The population model described in the previous
section thus determines all the parameters necessary to evaluate
whether a given binary system will be observed by, e.g., the LIGO
detector system.

First predictions for the chirp mass distribution of observed systems 
were recently obtained by \cite{bulik}, which considers varying underlying 
binary parameters.  Additionally, \cite{bulik2} studied the influence of 
different cosmological models on the observed chirp mass distribution.  
Complementing these studies, we determine the distribution of observed systems from the same physical BCO
population model, however we include the effects of anisotropic distribution
of galaxies while we do not consider the different underlying cosmological
assumptions. Our method is as follows.

We first draw a representative sample of binary systems populating each galaxy; we evaluate the expected signal-to-noise in the detector system; we determine the observed population by choosing the subset of systems whose signal-to-noise in the detector system is greater than the detection threshold.
Throughout this work we assume a signal-to-noise threshold for
detection $\rho_0$ equal to 8. The results presented in this work are
based on $\simeq\!10^6$ binaries drawn randomly from $\simeq\!75000$ galaxies. The ratio of the observed number of systems to the number of systems in the parent population defines the detection efficiency. The rate of detected inspiral events is equal to the merger rate in the physical population reduced by the detection efficiency. 

\subsection{Signal-to-Noise Ratio}
\label{sec:snr_properties}

The strength of a signal event observed in a detector is characterized
by the event's signal-to-noise ratio $\rho$.  The anticipated $\rho$ value associated with a particular source depends on
the detector, the source, and the method of data analysis. For
coalescing compact object binaries there is an optimal method of
analysis --- matched filtering --- that allows us to express the
{\it anticipated} $\rho$ for a particular binary system
in terms of its distance, component masses, and orientation with
respect to the detector. Even for identical sources the actual
signal-to-noise ratio will vary from instance to instance owing to the
stochastic character of detector noise.

The expected signal-to-noise $\rho$ associated with a compact object
binary observed in a single detector using the technique of matched
filtering was first derived by \cite{finn_chernoff}; here we use the
particular expression 
as given in \cite{finn96}:
\begin{subequations}\label{eq:snr}
\begin{equation}
\rho_D  \equiv 8 \Theta  \left( \frac{r_{0}}{D} \right) \left(
  \frac{\mathcal{M}}{1.2 M_{\odot}} \right)^{5/6} \zeta (f_{\max})    
\label{eq:A}
\end{equation}
where
\begin{eqnarray}
\mathcal{M} &=& \mu^{3/5}M_{tot}^{2/5}\label{eq:chirpMass},\\
r^2_{0} &\equiv& \frac{5}{192 \pi } \left( \frac{3}{20} \right)^{5/3} 
x_{7/3} M_{\odot}^2 \qquad,
\label{eq:r_0}\\
x_{7/3} &\equiv& \int_{0}^{\infty} \frac{df (\pi M_{\odot})^2}{(\pi f
  M_{\odot})^{7/3} S_h(f) },\\
\zeta(f_{\max}) &\equiv& \frac{1}{x_{7/3}} \int_{0}^{2f_{\max}}
  \frac{df (\pi M_{\odot})^2}{(\pi f M_{\odot})^{7/3} S_h(f)} ,
\end{eqnarray}
\end{subequations}
$D$ is the luminosity distance to the binary system, $\Theta$ is a
function of the relative orientation of the binary (including its
orbital plane inclination) and the detector, $M$ and $\mu$ are the
system's total and reduced mass, $S_h(f)$ is the detector noise power
spectral density in units of squared GW strain per Hz,
and $f_{\max}$ is the GW frequency at which the
inspiral detection template ends (and which is no greater than the
frequency at which the inspiral itself transitions into coalescence).

The LIGO detector system consists of not one but three detectors: a
4~Km interferometer (H1) and a 2~Km interferometer (H2) located near
Hanford, Washington, and a 4~Km interferometer (L1) located near
Livingston, Louisiana. The signal-to-noise associated with any given
inspiraling BCO will be different in the three interferometers, owing
to differences in their respective noise power spectral densities
$S_h$ and their geographic locations and orientations. When we speak
of the signal-to-noise associated with an compact object binary
inspiral observed in LIGO we will assume that these three
interferomters are used together as a single detector following
\cite{finn01}, in which case the signal-to-noise ratio for the network
of three detectors can be approximated, for LIGO, by the quadrature
sum of the signal-to-noise ratios in the individual detectors:
\begin{equation}
\label{network}
\rho^2 = \rho^2_{\mathrm{H1}} + \rho^2_{\mathrm{H2}} +
\rho^2_{\mathrm{L1}},
\end{equation}
where $\rho_{\mathrm{H1}}$, $\rho_{\mathrm{H2}}$ and
$\rho_{\mathrm{L1}}$ refer to the expected signal to noise of the
binary in the H1, H2 and L1 detectors, calculated via \ref{eq:A},
respectively. 

In the following subsections we discuss $\zeta$, $f_{\max}$ and
$\Theta$ in more detail. 

\subsection{$\zeta$ and $f_{\max}$}

The chirp mass $\mathcal{M}$ and frequency $f_{\max}$ are intrinsic
properties of the BCO. The GW signal from an
inspiraling binary system is nearly sinusoidal with a frequency twice
the binary's orbital frequency. Consequently, over time the signal
enters the detector's effective bandwidth from low-frequency, travels
across the band as time elapses, and --- if the frequency where the
inspiral ends and coalescence takes place is high enough --- leaves
the detector's bandwidth at high frequency. The quantity
$\zeta(f_{\max})$ is unity when the BCOs inspiral phase completely
covers the detectors effective bandwidth and is less than unity to the
degree that the inspiral terminates within the detectors band, or before the signal enters the band. 

We model $f_{\max}$ by assuming that the BCO inspiral phase proceeds
until an {\it innermost circular orbit} (ICO) is reached, at which
point the components coalesce in less than an orbital period.  Using
the post-Newtonian approximation together with a compelling ansatz
\cite{kww} have estimated the ICO orbital frequency for binaries with
components or equal mass to be
\begin{equation}
\label{fmax}
f_{\mathrm{ICO}}= 710 Hz \left( \frac{2.8 M_{\odot}}{M}   \right). 
\end{equation}
We assume that $f_{\max}$ is equal to $f_{\mathrm{ICO}}$. 

For binaries with unequal mass components, $f_{\mathrm{ICO}}$ depends
additionally on a function of $\mu/M$.  For the binaries that we
encounter in our Monte Carlo procedure (where the mass ratio of the
larger over the smaller is rarely above 2.5), the effect of the mass
asymmetry is small.  We ignore this small correction, making exclusive
use of eq.\,(\ref{fmax}).

The question of how the transition from inspiral to coalescence takes
place is far from settled. The \cite{kww} calculation of
$f_{\mathrm{ICO}}$ treats the binary components as point masses,
ignores hydrodynamic effects, and employs an ansatz that --- while
compelling --- is still a guess, based on an analogy to Schwarzschild
spacetime. A number of authors have discussed hydrodynamic effects
that may lead to coalescence at orbital frequencies less than
$f_{\mathrm{ICO}}$ \citep{faber,lai96,lai94}, and numerical
simulations of relativistic binary systems have led to other results
for the $f_{\mathrm{ICO}}$ orbital period
\citep{tani,grand02,cook,baum}. Future work will need to explore the
dependence of the observed rates and chirp mass distribution on this
uncertainty as well.

\subsubsection{The Antenna Projection $\Theta$}\label{sec:Theta}

Interferometric detectors like LIGO are sensitive to a single
GW polarization and have a quadrupole antenna pattern
\citep{thorne}. The function $\Theta$, defined in detail in
\cite{finn_chernoff,finn96} describes the dependence of the inspiral
signal-to-noise ratio on the position and orientation of the binary
relative to the detector. It ranges from 0 to 4, with typical values
near unity {\it when we consider an isotropic distribution of
  sources.} In the local neighborhood, however, the distribution of
galaxies is not isotropic on the detector's sky: e.g., there are
significant concentrations of galaxies in the direction of the Virgo
and the Fornax clusters, which are at a fixed declination relative to
the LIGO detectors. Since the signal-to-noise ratio of a binary is
directly proportional to $\Theta$, binaries from galaxies at some
declinations are more likely to be detected than binaries in others
and this plays an important role in relating the observed event rate
to the underlying, physical event rate.

For example, Figure \ref{fig:Theta_dist} compares the distribution of
$\Theta$ for binaries associated with Virgo cluster galaxies
(solid line) as observed at the LIGO Hanford detector to the
distribution of $\Theta$ for binaries isotropically distributed about
the sky. Note that Virgo is relatively poorly positioned in
declination relative to the LIGO Hanford detector, leading to a
smaller number of sources with large $\Theta$ then we would expect for
an isotropic distribution of binaries. 

\subsection{The noise power spectral density $S_h(f)$}
\label{sec:detector_prop}

The influence of the detector on the signal-to-noise ratio $\rho$
enters through the quantity $r_0$ and the function $\zeta(f)$, each of
which depend on the detector's strain-equivalent noise power spectral
density $S_h(f)$ (cf.\ eq.\ \ref{eq:snr}). 

A GW signal incident on an interferometric detector
like LIGO leads to a response that can be characterized as a
projection $h(t)$ of the incident GW strain acting on
the detector arms. Measurement noise, which is contributed at many
points in the transduction chain, is indistinguishable from a
GW signal $h_n(t)$ and we characterize the
{\it strain-equivalent noise} by its mean square amplitude,
$\overline{h_n^2}$. The strain-equivalent noise power spectral density
$S_h(f)$ is the contribution to $\overline{h_n^2}$ from noise
components in a 1~Hz bandwidth about the frequency $f$, so that
\begin{equation}
\overline{h_n^2} = \int_0^\infty df\,S_h(f).
\end{equation}

LIGO is currently in the late stages of commissioning, with results
from early data sets under review \citep{ligoR,pulsar,inspiral}.  When
commissioning is complete the noise power spectal density $S_h(f)$
will meet or exceed the specification given in \cite[3-2]{SRD}. To
evaluate the sensitivity of the initial LIGO instrumentation we use a
parameterized version of this noise curve \citep{Owen},
\begin{eqnarray}\label{eq:srd}
S_h\left(f\right) &=& 9\times 10^{-46}\,[\left(4.49 x\right)^{-56}
 + 0.16 x^{-4.52} + 0.52 + 0.32 x^2] \,{\mathrm{Hz}}^{-1}\\
 x &\equiv& \frac{f}{150 {\mathrm{Hz}}}, 
\end{eqnarray}
for each of the two LIGO 4~Km interferometers (referred to as H1 and
L1). For the 2~Km  Hanford detector (referred to as H2) we approximate
$S_{h,2km}(f)$ as $4\,S_{h(f)}$.\footnote{Not all noise sources scale
  with length in a simple manner: in particular, the laser shot noise
  spectrum depends on the Fabrey-Perot arm cavities in a more
  complicated manner. Nevetheless this approximation is more than
  suitable for our purpose here.} As a consequence of this
approximation note that for the initial LIGO detectors
\begin{itemize}
\item $\zeta(f_{\max})$ is the same function of $\mathcal{M}$
for H1, H2 and L1 approximation; and 
\item $r_0$ for H1 and L1 are identical and twice the appropriate for
  H2:
\begin{equation}
r_{\mathrm{H1}} = r_{\mathrm{L1}} = 2r_{\mathrm{H2}} = 7.7
\end{equation}
\end{itemize}

The LIGO Laboratory and the LIGO Scienctific Collaboration have also
recently proposed an advanced set of LIGO detectors. In addition to
extending the 2~Km Hanford interferometer to a full 4~Km these
advanced detectors have a much lower detector noise and a greater
effective bandwidth for BCO inspiral searches. We have also evaluated
the event rates and distributions using the current estimates for the
limiting noise sources associated with this advanced detector system (Shoemaker 2003, private communication). Figure \ref{fig:noise} shows the target noise curve
for the initial LIGO detectors (eq.[\ref{eq:srd}]) and the current
estimate for the advanced LIGO limiting noise curve.  Again note that,
as a consequence of this approximation
\begin{itemize}
\item $\zeta(f_{\max})$ is the same function of $\mathcal{M}$
for H1, H2 and L1 approximation; and 
\item $r_0$ for H1, H2 and  L1 are identical for the advanced LIGO
  detectors:
\begin{equation}
r_{\mathrm{H1}} = r_{\mathrm{L1}} = r_{\mathrm{H2}} \simeq 120
\end{equation}
\end{itemize}

\section{RESULTS} 
\label{sec:results}

In this section we discuss how the differential sensitivity of the
LIGO detectors, as a function of source sky position, chirp mass,
distance and analysis threshold, can be combined with the spatial and chirp
mass distribution of coalescing binaries, to determine how the
observed coalescing binary distribution reflects the underlying,
physical distribution. 

\subsection{Binary Compact Object Populations in Galaxies}

In addition to its dependence on distance, sky position and orbital
plane inclination, the sensitivity of GW detectors
like LIGO to a particular BCO depends on a function of the binary's
component masses. We use the BKB binary synthesis
calculations to populate each galaxy with a distribution of binary
systems of different chirp mass\footnote{For the most part of \S\ref{sec:results}, we confine our attention to the reference model A from BKB. However we explore the full set of BKB models in  \S\ref{sec:cum}}. The
luminosity scaling of~\cite{phinney} is naturally applicable to our
approach and we set the total inspiral rate for a particular galaxy
equal to the Milky Way rate $R_{\mathrm{det,MW}}$ times its blue
luminosity (corrected for reddening) in units of the Milky Way blue
luminosity. Note that $R_{\mathrm{det,MW}}=R_{\mathrm{MW}}$, since the initial LIGO detectors are expected to detect essentially all BCO inspirals in the Milky Way with chirp mass less than 18~$\mathrm{M}_\odot$ (that exceeds the typical expected maximum chirp mass; see Figure 2).

\begin{equation}
\mathcal{R}_{\mathrm{gal}} = \frac{L_{\mathrm{gal,b}}}{L_{\mathrm{MW}}}
R_{\mathrm{MW}}.
\end{equation}

Our principal tool for understanding how the physical distribution of
coalescing binaries is reflected in the observed distribution is the
detection efficiency. We define the detection efficiency as the
fraction of binaries from a given population that are detected with a signal-to-noise ratio higher than a chosen threshold (in this study: 8). We may
thus consider the detection efficiency for all binaries in a given
galaxy, or the efficiency for NS-NS binaries over all galaxies, or the
efficiency for detection of binaries at a given sidereal time, etc.

Under the assumption of uniform galaxy distribution, the natural
measure of a detector's sensitivity is the volume of space surveyed,
which is conveniently expressed as an effective radius $r$ such that
the surveyed volume is equal to the volume of a sphere of radius $r$
\citep{finn_proceed}. When taking into account the actual galaxy
distribution and the variation in binary populations among galaxies a
different measure of sensitivity suggests itself: the effective number
of Milky Way galaxies surveyed,
\begin{equation}
N_{G} \equiv \frac{R_{\mathrm{det}}}{R_{\mathrm{MW}}} =
\sum_{i}{\frac{L_{i}}{L_{\mathrm{MW}}}}f_{\mathrm{det,i}}, 
\end{equation}
where $f_{\mathrm{det,i}}$ is the detection efficiency and $L_i$ is the blue
luminosity for galaxy $i$, $L_{\mathrm{MW}}$ is the Milky
Way's blue luminosity, and we have used the blue luminosity scaling
$L_{i}/L_{\mathrm{MW}}$ as discussed in \S\,\ref{sec:approach}. The concept of $N_G$ is more appropriate for our approach, since it takes into account the actual galaxy distribution and the BCO mass distribution. As detectors become more sensitive, $N_G$ grows accordingly. 

In this formulation $R_{\mathrm{det}}=N_{G}\,R_{\mathrm{MW}}$. The Galactic inspiral rate $R_{\mathrm{MW}}$ can be calculated based on the current observed sample for NS-NS \citep{kal03} and from population synthesis calculations (e.g., BKB).

\subsection{Detection efficiency and sidereal time}

Interferometric detectors like LIGO are most sensitive to
gravitational waves of a single polarization incident normal to the
plane of the detector's arms. As the Earth rotates about its axis, the sky
locations and polarization of sources to which it is most sensitive
rotate as well. Since galaxies are not uniformly distributed about the
sky the expected rate of detected coalescence events is periodic with
the sidereal day. The detailed variation depends on the geographical
distribution of galaxies, the distribution of coalescing binaries with
system chirp mass, and the signal-to-noise threshold for detection.
Figure \ref{fig:time} shows the probability density for detections
under the assumptions that the geographical distribution of galaxies
is determined by our galaxy catalog approach (cf.\ \S\ref{sec:distribution}), the
chirp mass distribution of coalesinc binaries is given by BKB model
A, and the signal-to-noise ratio $\rho_0$ threshold for detection in
initial LIGO is $8$. Here time $t$ is measured in hours and is given
by UT plus the sidereal time at Greenwich at 0h UT.  The Virgo Cluster
is above a point in between the Hanford and Livingston detectors for
approximately the duration $18 h< t < 20 h$, closely coinciding with
the peak in Figure\,\ref{fig:time}.

\subsection{Dependence of detection range and efficiency on declination}

Setting aside the actual distribution of galaxies, it is interesting to
note the dependence of the detectors ``range'' --- or distance to the
most distant observable source above threshold --- on source sky
position. When sampled over sidereal time, the range depends on the
sky position only through declination.  Figure \ref{fig:D_opt} shows,
for initial LIGO, the detector range as a function of declination,
normalized to $1.4+1.4 ~\mathrm{M}_{\odot}$ NS-NS binaries. For an advanced LIGO 
with the H2 interferometer extended to 4~Km the declinations of maximum 
range will shift slightly toward the zenith and nadir of L1.

Figure \ref{fig:f_dec} shows the dependence of detection efficiency
--- i.e., the fraction of binaries detected --- on declination for
galaxies at the distance of the Virgo cluster.  Note that this dependence on declination will change if the ground-based detectors are located differently.  In particular, the sensitivity of detectors to inspirals from the Virgo cluster improves for detectors with lattitudes corresponding to declinations near that of Virgo.  Indeed, Livingston and VIRGO, due to the relative proximity of their latitude and the Virgo cluster's declination, are more optimally placed relative to the Virgo cluster than the Hanford or GEO detectors.  Interestingly, had H1 and H2 been constructed exactly as they are, but instead with the present lattitude of the VIRGO detector, the LIGO network would detect about 25\% more NS-NS inspirals (at design sensitivity) from the Virgo cluster.

Note that the maximum detection efficiency at this distance in Figure \ref{fig:f_dec} occurs for galaxies located at
the celestial poles. This might be surprising since the detectors are
most sensitive to sources at their zenith or nadir. The reason for
this apparently paradoxical result is that a galaxy at a declination
corresponding to the detector latitude is only at the detector's zenith
for a short fraction of a sidereal day, while the efficiency for
galaxies at a celestial pole are independent of sidereal time. Despite
the shorter duty cycle for sources at the detector zenith or nadir, it
is still the case that galaxies at these locations will be seen to
greater distances.

\subsection{Dependence of detection efficiency on both galaxy and BCO mass distributions}

Figure \ref{fig:4_panel} shows the detection efficiency of every
galaxy in our catalog, plotted as a function of galaxy distance. The
top panel shows the total efficiency for the complete population of
binaries. Subsequent panels shows the efficiencies for the BH-BH,
NS-BH and NS-NS sub-populations for initial LIGO.  Galaxies at a given
distance have a range of efficiencies owing to their different
declinations. This may amount to as much as a 40\% variation.
  
Previous studies adopted a detection efficiency for initial LIGO of
unity up to distances of 20 Mpc for NS-NS, 40 Mpc for NS-BH, 100 Mpc
for BH-BH, and 46 Mpc for the entire population, and zero beyond.
These cut-offs are shown as vertical dashed lines. In our more
realistic approach one sees non-zero efficiencies to much greater
distances. For example BH-BH binaries are observed with 10\%
efficiency even at distances of 130 Mpc. On the other hand, inside the
previously adopted cut-off distances the efficiency is not 100\%: for
example, the overall efficiency to Virgo cluster inspirals is 60\%.
  
The NS-NS, NS-BH and BH-BH sub-population efficiencies vary
differently with distance owing to their different chirp mass
distributions. Figure 9 shows the expected distribution
with chirp mass of detected binaries. This should be contrasted with
the actual population distribution, as shown in Figure 2. As expected 
the distribution is dominated by
BH-BH binaries; however, the relative proportion of detected NS-NS
binaries is more than double the proportion previously calculated
assuming a uniform galaxy distribution \citep{bulik}.  The increased
proportion of NS-NS relative to BH-BH binaries arises because the
uniform galaxy distribution assumption significantly underestimates
the number of galaxies in the nearby universe. As is apparent from
Figure 1, this underestimate is greater at Virgo
cluster distances, where there is still significant efficiency for
NS-NS binaries, than at, e.g., 100 Mpc where only BH-BH are detected.
Correspondingly, the boost in the number of NS-NS binaries, relative
to the uniform distribution model, is greater for NS-NS binaries than
it is for BH-NS binaries and for BH-BH binaries. (The sensitivity of
the advanced LIGO detectors is sufficiently great that the uniform
assumption is applicable and the distribution of detected sources will
follow \cite{bulik} more closely.) From Figure 9 we also note that about half of the detected binaries are expected to have chirp mass in the range $5-9$\,M$_\odot$, with a peak at 9\,M$_\odot$ (corresponding to 2 10\,M$_\odot$ black holes). These results can be folded with the most current Galactic rate estimates to estimate the number of black hole inspiral events with chirp masses in the above range that we might expect from initial LIGO observations in one year at full sensitivity. It is clear that data searches would need to focus on these higher masses, where however the GW waveforms are harder to calculate. Nevertheless the necessary effort with focus on these systems is needed, since  limits in  this range will be most constraining to population models.

\subsection{Effective number of MWEG surveyed}
\label{sec:cum}

The naive use of an effective radius can lead to a misunderstanding in
the overall rate and nature of detected coalescing binary systems.
Figure 10 shows the cumulative rate of detections by
the initial LIGO detector from sources within a distance $D$. The
dashed line shows the cumulative rate assuming a uniform distribution
of galaxies and a 100\% efficiency for binaries within a distance $r$,
with $r$ equal to 20, 40, and 100 Mpc, for NS-NS, NS-BH, and BH-BH
binaries respectively.  The solid line show the approach taken here:
i.e., the rate of detected binaries within a distance $D$ taking into
account the actual galaxy distribution, the estimated number of
binaries per galaxy, the distribution of binaries with chirp mass, and
the overall detection efficiency for each galaxy.  The solid lines
show evidence that particular geographic features in the distribution of
galaxies have sharp effects on rate of detections. The most striking
example is the Virgo Cluster (the spike at 20 Mpc), which comprises,
for example, 20\% of NS-NS detections.  In general, NS-NS
extrapolation factors based on the uniform galaxy density assumption
have been underestimated by a factor of $\simeq 2-3$, while BH-BH
extrapolation factors have been slightly overestimated by a factor
$<100\%$.

Even though the quantity $N_{G}$ is {\it normalized} to the Galactic
rate, it still depends on the assumed model chirp mass distribution.
To examine this dependence we consider all models calculated in BKB
that are considered realistic at present. We calculate the associated
$N_{G}$ values for each of the three BCO populations shown in Figure
11. These models differ in terms of a number of
parameters that determine single-star and binary evolution (e.g.,
stellar winds, common envelope evolution, NS and BH kicks, etc; for a
detailed discussion see BKB). Based on these models it
is evident that the variation in $N_{G}$ is lower than a factor of
1.5-2 for the three types of BCO populations. We consider these 
variations to represent the systematic uncertainty associated with the
extrapolation factor $N_{G}$.

Finally, we explore the growth of $N_{G}$ with detector sensitivity.
As LIGO's sensitivity improves, approaching design sensitivity for
initial and advanced detectors, of particular interest is the growth
of $N_{G}$ for NS-NS binaries. In lieu of a detection, $N_{G}$ is
important for assessing upper rate limits for NS-NS inspiral in the
Milky Way. A measure of the sensitivity improvement is the increase of
the value of $r_0$ (see eq.\,\ref{eq:r_0}) or the ratio
\begin{equation}
\xi \equiv \frac{r_{0}}{r_{0,SRD}},
\end{equation}
where $r_{0,SRD}$ is calculated similarly but with the advanced LIGO
noise curve.  In Figure 12 we show the dependence
of $N_{G}$ on $\xi$ for the initial LIGO configuration of 4 and 2 km
detectors in Hanford, Washington, and a 4 km detector in Livingston,
Louisiana. The effective number of MWEG surveyed increases as $x^{P}$,
initially with $P<1$, but growing with $P\simeq 2.6$ as design sensitivity is reached.
This rate represents an increase by a factor of
$\simeq 300$ Milky Way equivalent galaxies compared to the sensitivity
of the detectors during the first Science Run~\citep{ligoR}.

\section{CONCLUSIONS}
\label{sec:conclusions}

In anticipation of the development of GW astrophysics
in the next several years, we consider the effects of observational
selection effects on the detectability of BCO inspiral events. Our
primary goal is to develop a realistic framework for the astrophysical
interpretation of rate constraints (from upper limits or inspiral
detections) anticipated in the next few years. This interpretation
should account for the main selection effects associated with
ground-based GW observations and properly constrain
models of radio pulsar and BCO populations. As a result of our
calculations we also make realistic estimates for the extrapolation of Galactic inspiral detection rates
based on the known spatial distribution of galaxies in the nearby
universe and the expected mass distributions of binary compact
objects.

Our results are summarized as follows:
\begin{itemize}
  
\item The local distribution of galaxies mostly relevant to NS-NS
  detections with initial LIGO is in fact very different from
  isotropic in sky direction and volume density. Most importantly the
  Virgo cluster represents a significant step in the cumulative blue
  luminosity (or the cumulative number of MWEG) all concentrated at a
  given (rather unfortunate) sky position. Failure to properly account
  for this local distribution of BCO sources would lead us to
  underestimating the importance of an upper limit on the inspiral rate derived from GW observations.
  
\item Until this study, because of the assumption of isotropic distribution of
  galaxies, detection rates of NS-NS inspirals have been {\em
    under}estimated by factors of $2-4$ and BH-BH inspirals have
  been {\it over}estimated by nearly a factor of $2$. These factors include
  the systematic uncertainties due to the chirp-mass distributions
  that are not very well constrained.
  
\item Detections of inspiral events and measurements of compact object
  masses are expected to provide us with tighter constraints on the
  BCO mass distributions and thus on the physics of BCO formation.
  However, our analysis shows that mass distributions of detected BCOs
  are strongly skewed towards higher masses (because of their stronger
  signals) compared to the parent mass distribution (see Figures \ref{fig:chirpdist} and
  \ref{fig:chirp}). For our reference population model, we find that about half of detected inspirals correspond to binaries with high chirp masses ($5-9$\,M$_{\odot}$). Since event rate limits in this range will be most constraining to BCO models, it is evident that there is a need for the development of efficient search methods for such massive systems. Understanding the systematics of this bias will be crucial for the astrophysical interpretation of such detections.
  
\item Inspiral detection efficiency depends strongly on the host galaxy
  sky position and the binary orbit orientation with respect to the
  detectors; as a result the true maximum distance for an optimally
  oriented binary can exceed the average detection distance by more
  than a factor of two.

\item Using the current most favorable (at peak probability) estimates of NS-NS inspiral rates for the Galaxy \citep{kal03} and our results on $N_{G}$ values, we find expected initial LIGO detection rates in the range of one event per 200 -- 3 years (for the reference pulsar population model at 95\% confidence level the range is one event per 3 - 50 years).

\end{itemize}
 
From the various galaxy physical properties we have considered the
blue luminosity (corrected for reddening), but we have ignored galaxy
metallicity and star formation history. Both of these factors affect
the expected mass distribution of compact object binaries as well as
their birth rate, for a given luminosity. For example, metallicity
affects massive stellar winds and the final compact object masses.
This effect has already been taken into account in \cite{inspiral} where the Magellanic clouds have been reached by LIGO. In principle we would like to include these effects in our calculations (to the extent of our current understanding of binary evolution and how it is affected by these factors); however at present it does not seem possible since this information is not available in detail for every galaxy in the catalogs, and we chose to ignore these factors instead of include them for only a very small subset of sample.

With the work presented here we also advance a paradigm for using 
initial LIGO binary inspiral to constrain models of binary evolution 
and BCO formation and of pulsar population properties. Using the 
current estimates of NS-NS inspiral rates in MWEG \citep{kal03} and 
scaling to the BH-BH population, it is clear that LIGO should 
eventually provide an astrophysically significant bound on the rate of 
BH-BH inspirals in the nearby universe. In the context of a particular 
binary synthesis model, such a bound can be translated to a bound on 
the MWEG BH-BH inspiral rate as well as on the rate fir the NS-NS and 
NS-BH sub-populations. The derived bound on the NS-NS sub-population 
can be compared to the estimates that arise from binary pulsar 
observations \citep{kal03}. Note that GW observations may also directly 
bound the coalescence rate for the NS-BH and NS-NS sub-populations at a 
significant level. All these bounds will be consistent only for certain 
binary formation and synthesis models. In this way, GW observations 
will contribute to our understanding of compact binary formation and 
evolution.

\acknowledgments This work is partially supported by National Science
Foundation awards PHY 01-21420 (VK), PHY 00-9959 (LSF), and PHY
01-14375 (LSF, PN and VK), and a David and Lucile Packard Science \&
Engineering Fellowship (VK).  KB acknowledges support of the grant PBZ-KBN-054/P03/2001. 
PN and VK are also grateful for the warm
hospitality of the Center for Gravitational Wave Physics at Penn
State. The Center for Gravitational Wave Physics is funded by the NSF
under cooperative agreement PHY 01-14375.

\clearpage




\centerline{\psfig{figure=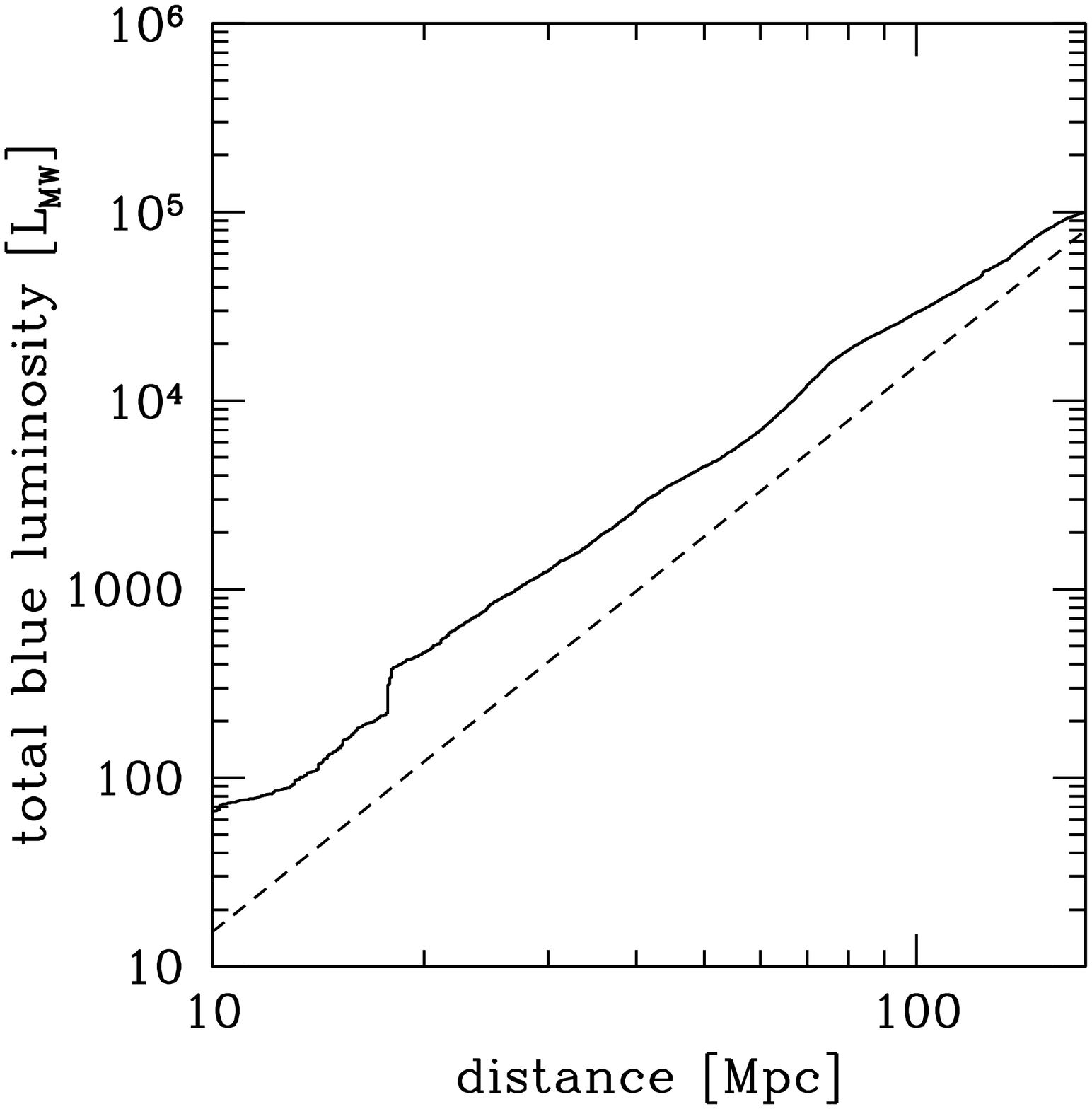,angle=0,width=5.2in}} 
 \figcaption{Cumulative blue luminosity enclosed within a distance
 $D$ using the LEDA galaxy catalog and correcting for incompleteness.
 The dashed line is produced assuming a uniform distribution of
 luminosity in space (using the normalization density adopted in
 KNST).  
 \label{fig:lum_cum}} 

\centerline{\psfig{figure=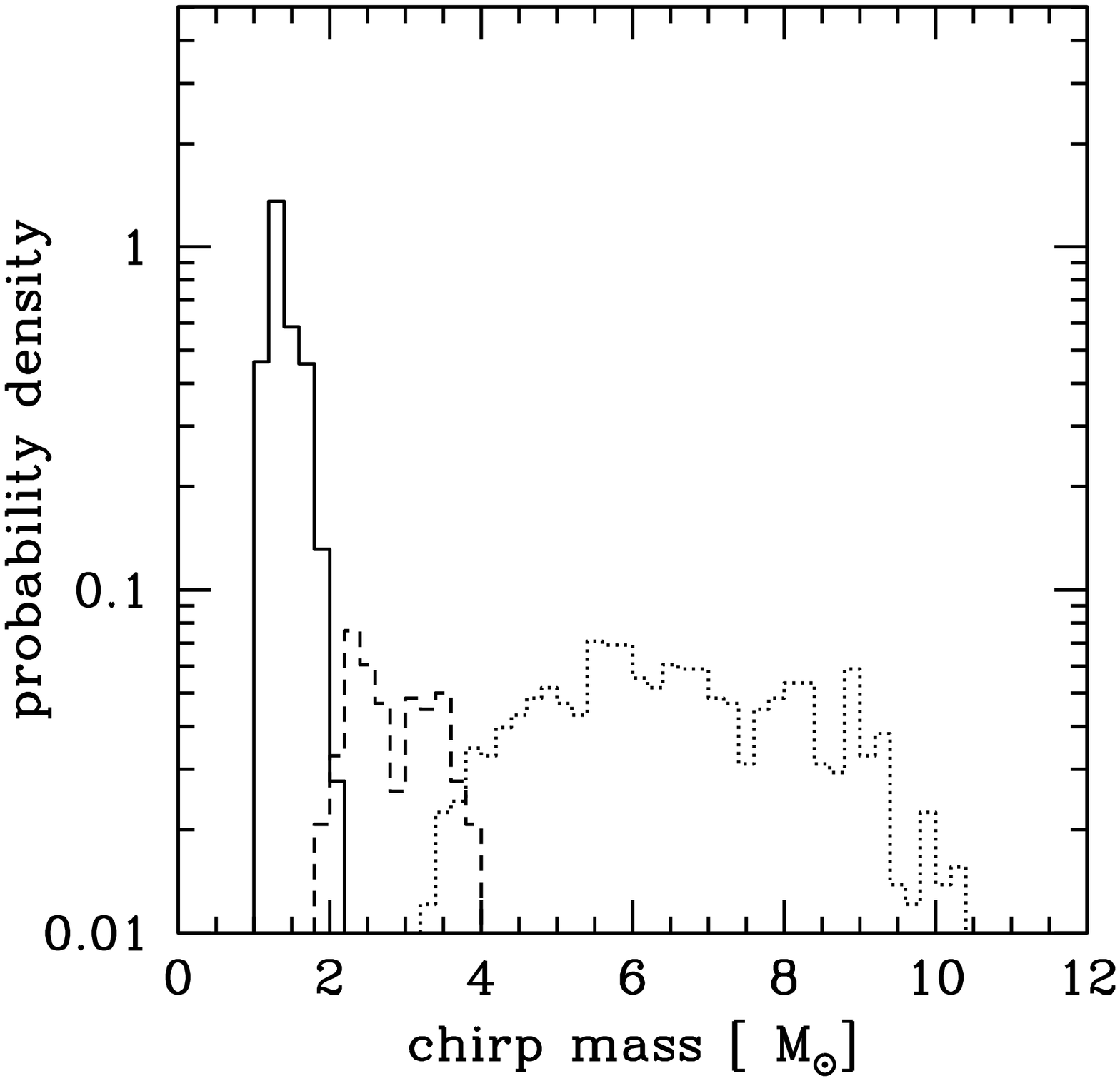,angle=0,width=5.2in}} 
\figcaption{The
  chirp mass distribution for inspiraling binaries, using model A from
  BKB, NS-NS (solid), NS-BH (dashed), and BH-BH (dotted)
  binaries. 
\label{fig:chirpdist}}

\centerline{\psfig{figure=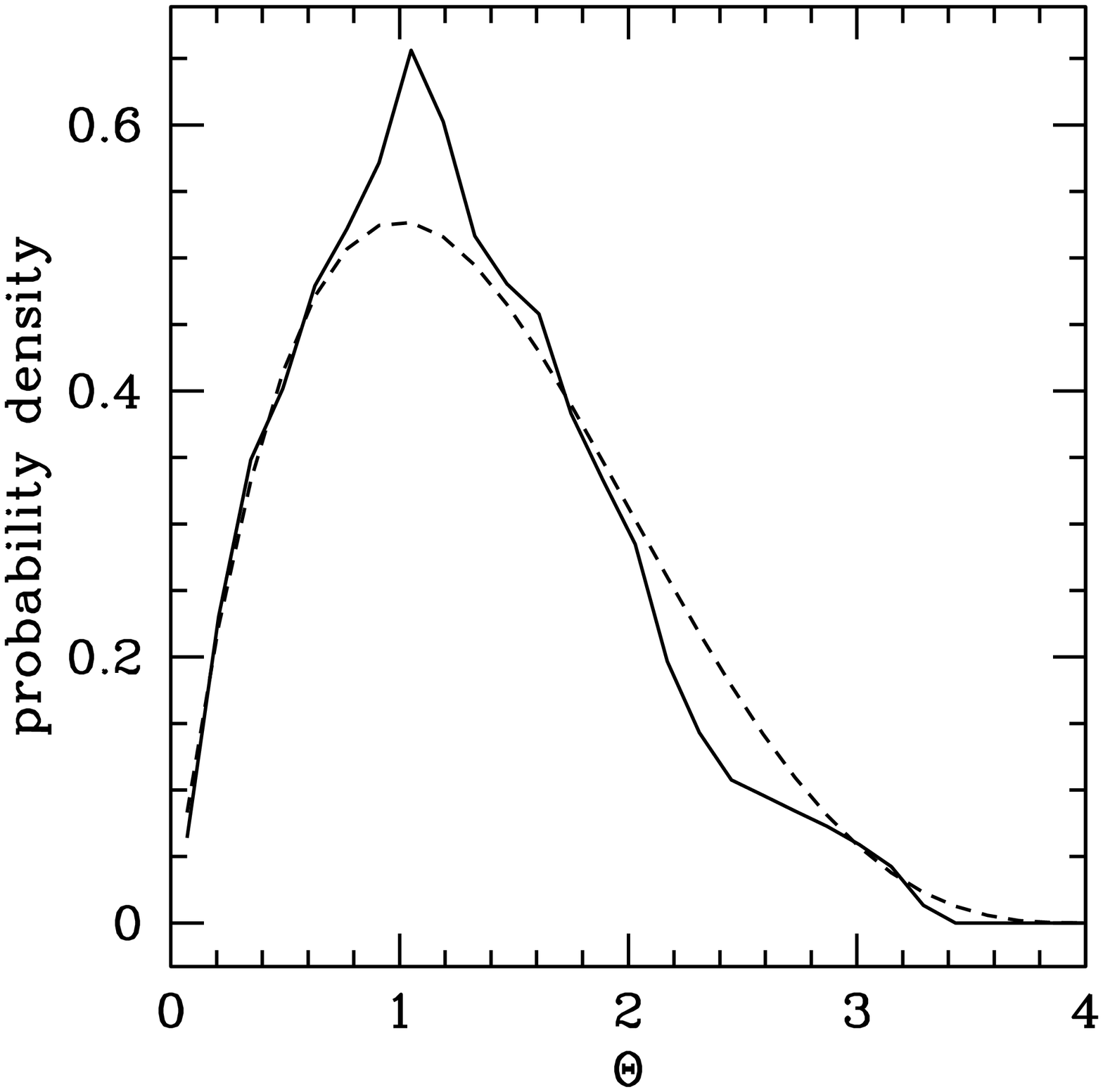,angle=0,width=5.2in}} 
\figcaption{The
  normalized probability distribution of $\Theta$.  The solid line gives the distribution for
  inspirals from galaxies in the Virgo Cluster relative to LIGO
  Hanford.  The dashed line gives the distribution assuming sources
  are uniformly distributed in space.
\label{fig:Theta_dist}}


\centerline{\psfig{figure=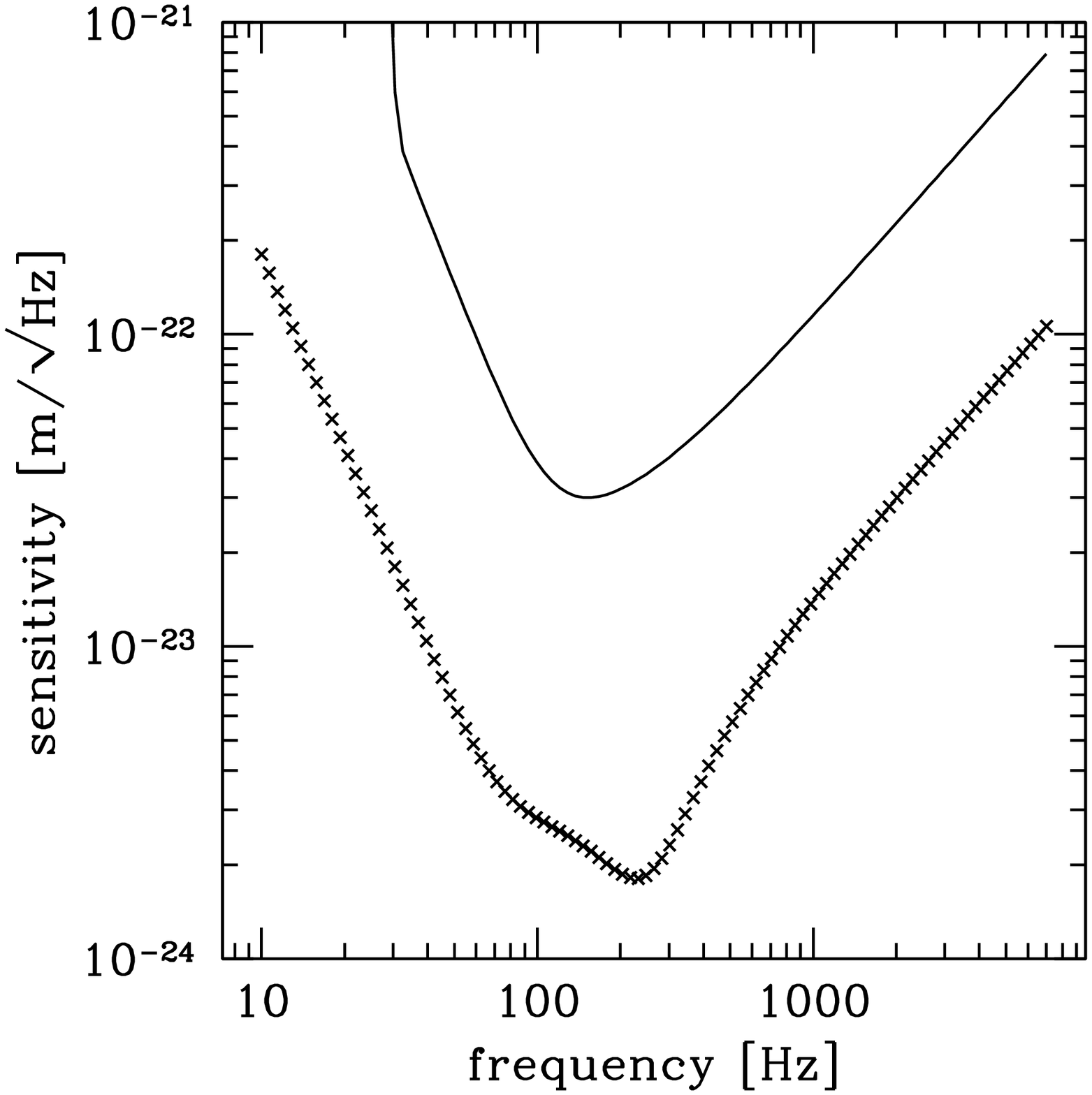,angle=0,width=5.2in}} 
\figcaption{Target noise curves for Initial (solid line) and Advanced (crosses) LIGO detectors.
\label{fig:noise}}

\centerline{\psfig{figure=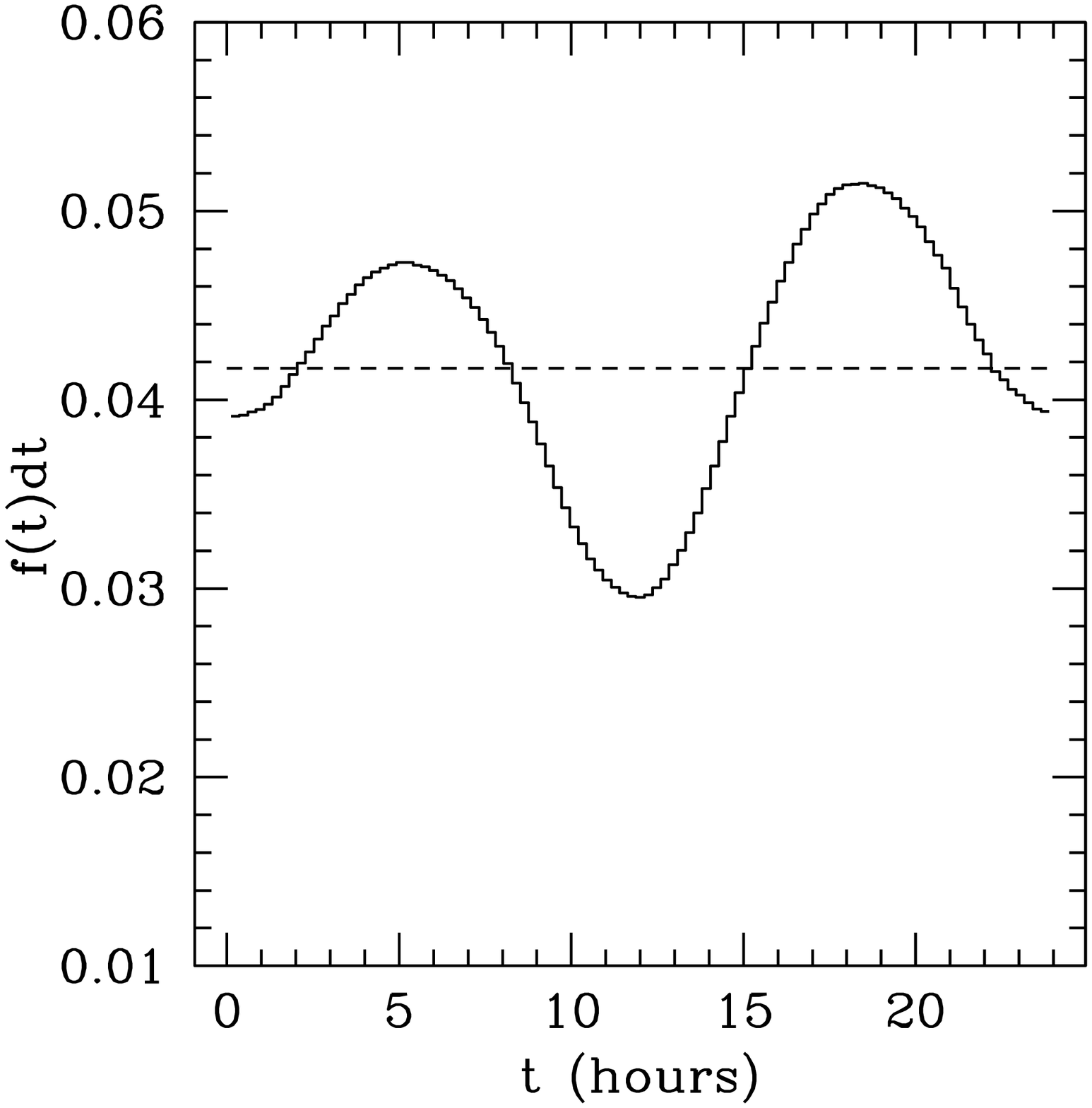,angle=0,width=5.2in}} 
 \figcaption{Probability $f(t) dt$ that an initial LIGO detection
 would come at an hour $t$ during the day. $t$ is the UT plus the
 sidereal time at Greenwich at 0 h UT.   The probability assuming
 uniform spatial distribution of galaxies is given for comparison
 (dashed line). 
 \label{fig:time}}


\centerline{\psfig{figure=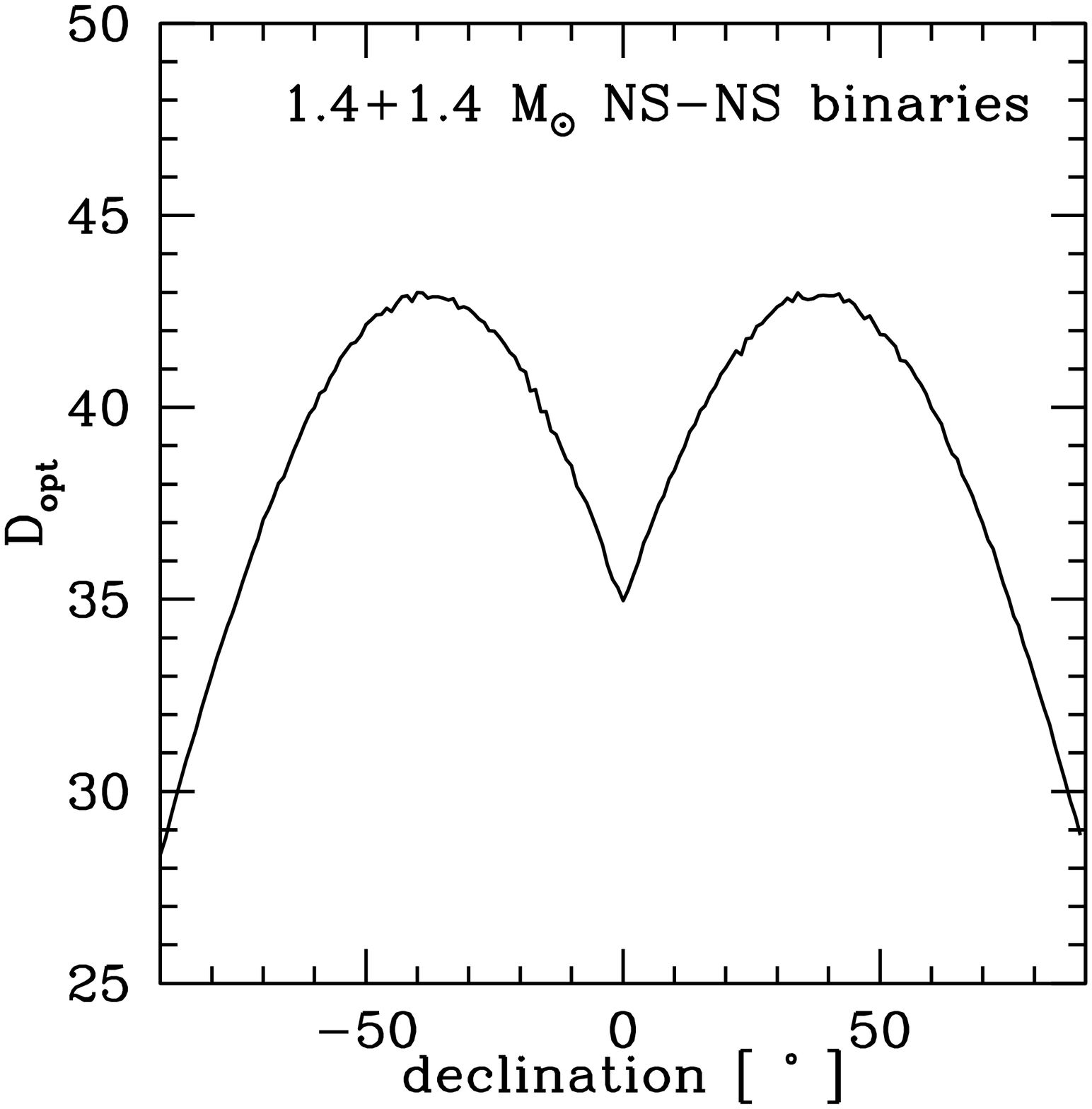,angle=0,width=5.2in}} 
 \figcaption{Maximum detection distance for the most optimally
 positioned binary as a function of declination.   LIGO Hanford and
 LIGO Livingston have lattitudes of $\simeq 45^{\circ}$ and $\simeq
 30^{\circ}$ respectively.  The peak in this plot corresponds to
 sources overhead a lattitude in between the detectors, weighted more
 towards Hanford due to its two interferometers versus Livingston's
 one. 
\label{fig:D_opt}}

 \centerline{\psfig{figure=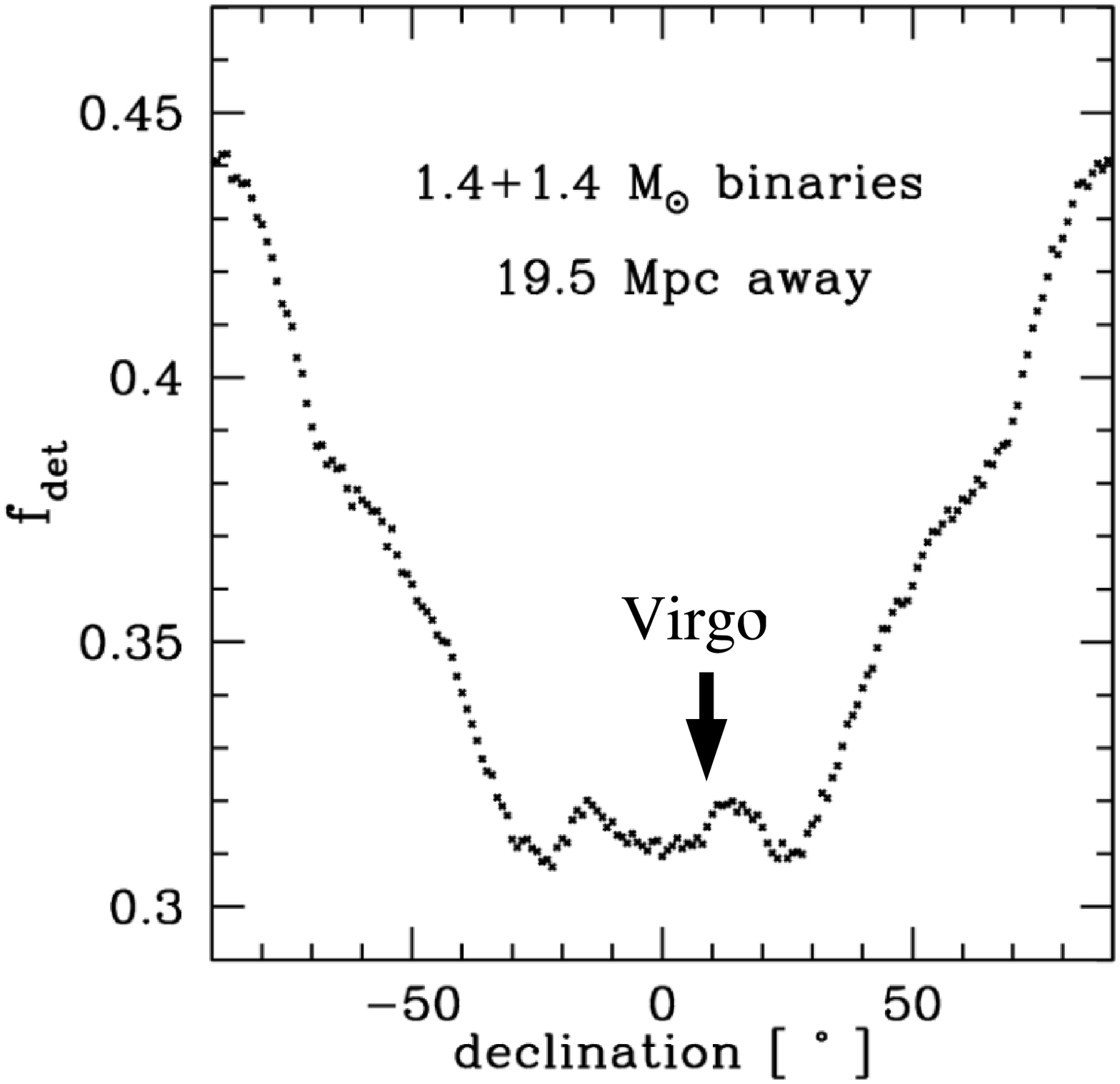,angle=0,width=5.2in}} 
 \figcaption{Detection efficiency as a function of declination, for
 $1.4+1.4 ~\mathrm{M}_{\odot}$ NS-NS binaries.  $f_{\mathrm{det}}$ is attenuated in
 regions near the equatorial plane, including the Virgo Cluster, 
which is a major source of NS-NS detections. 
\label{fig:f_dec}}

\centerline{\psfig{figure=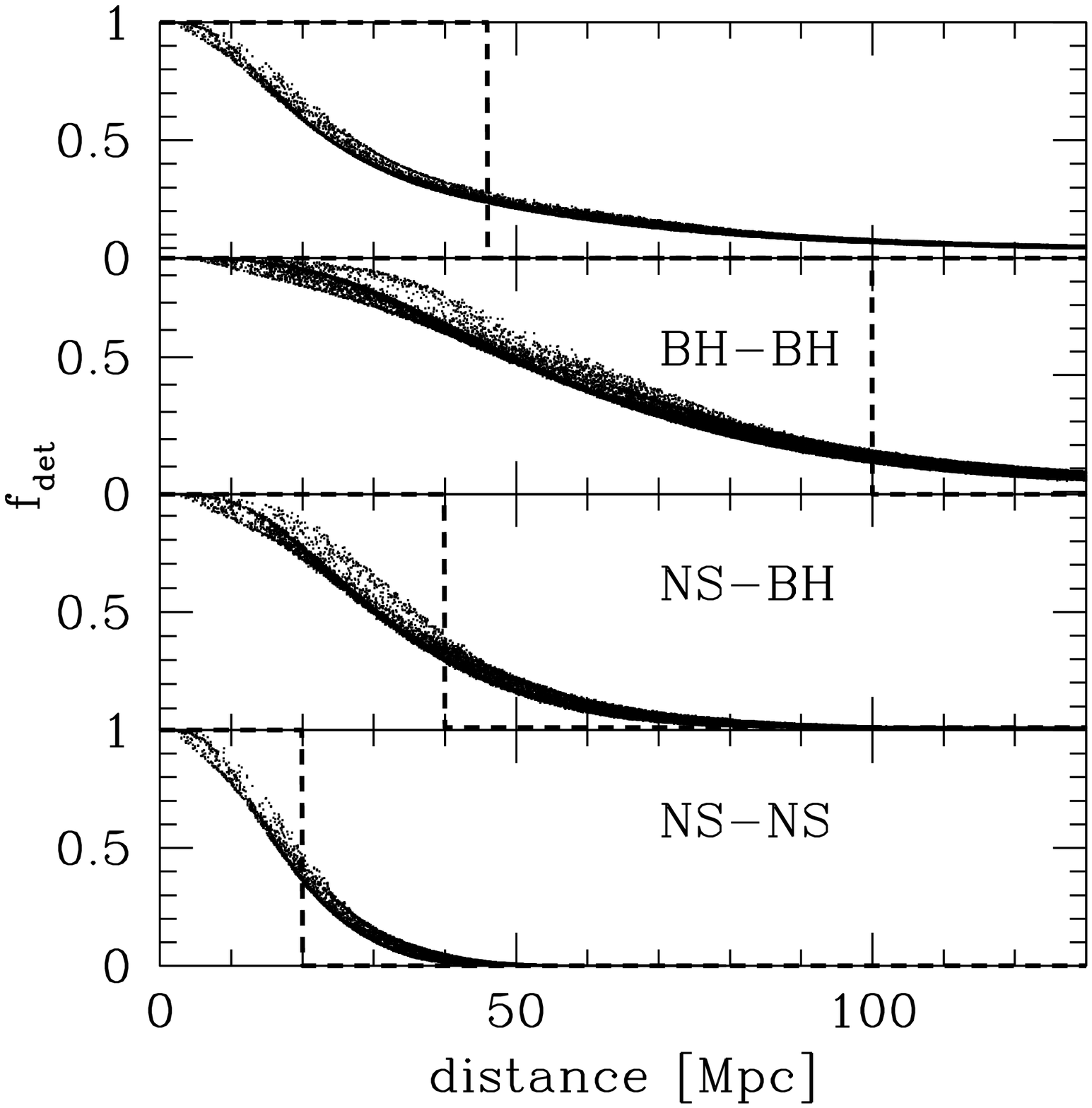,angle=0,width=5.2in}} \figcaption{
  The detection efficiency of every galaxy in our catalog, plotted as
  a function of galaxy distance. The top panel shows the total
  efficiency for the complete population of binaries. Subsequent
  panels shows the efficiencies for the BH-BH, NS-BH and NS-NS
  sub-populations for initial LIGO. Previous studies adopted a step-like detection efficiency for
  initial LIGO at 20 Mpc for NS-NS, 40 Mpc for NS-BH, 100 Mpc for
  BH-BH, and 46 Mpc for the entire population: these cut-offs are
  shown as vertical dashed lines.  
\label{fig:4_panel}}


\centerline{\psfig{figure=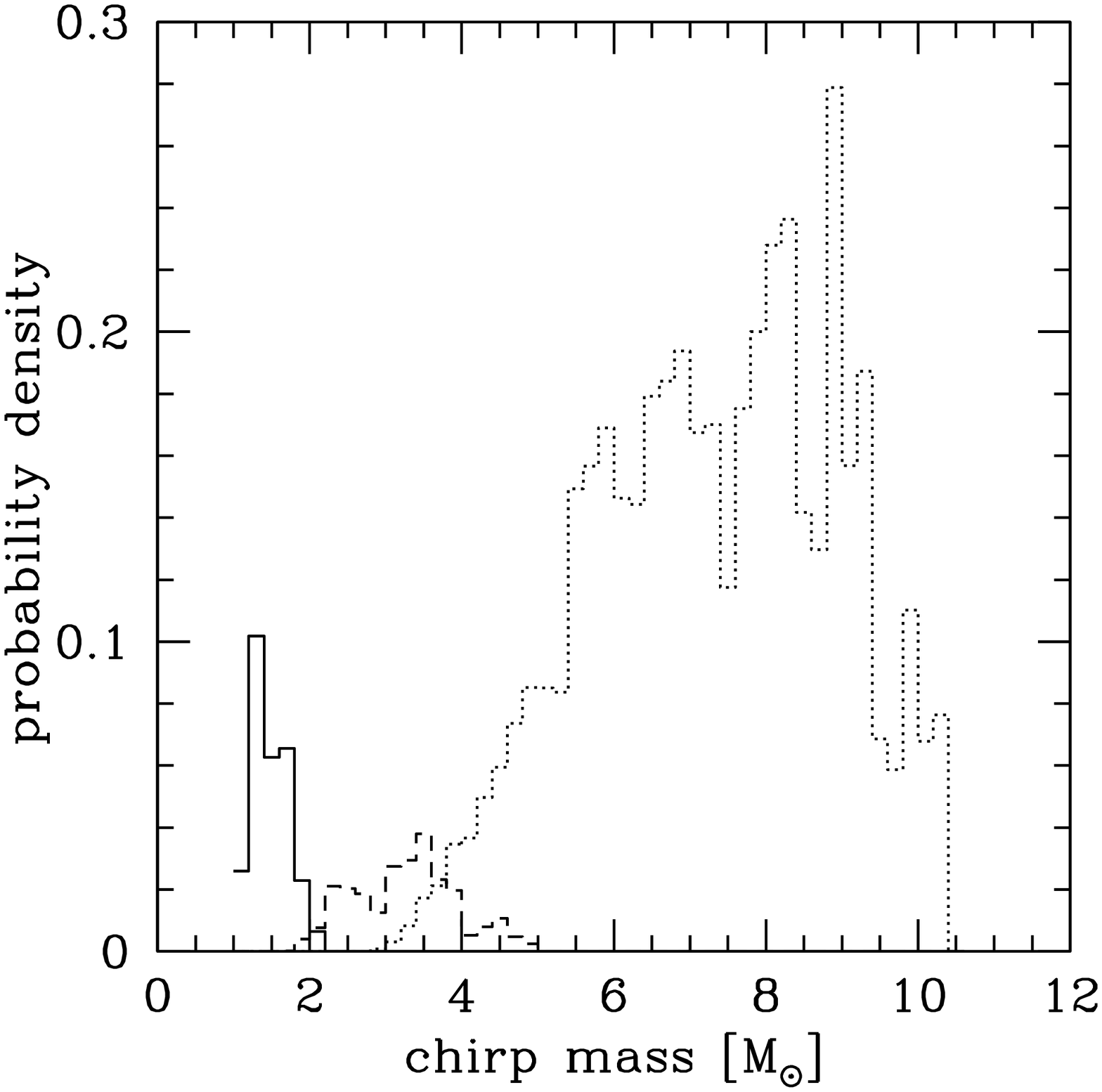,angle=0,width=5.2in}}
\figcaption{The expected normalized chirp mass distribution for
  detected inspirals, using model A from BKB.  The solid line gives
  the NS-NS binaries, the dashed gives NS-BH binaries, the dotted
  gives BH-BH binaries. Compare to the intrinsic chirp mass distribution in
  Figure\,\ref{fig:chirpdist}. 
 \label{fig:chirp}}


\centerline{\psfig{figure=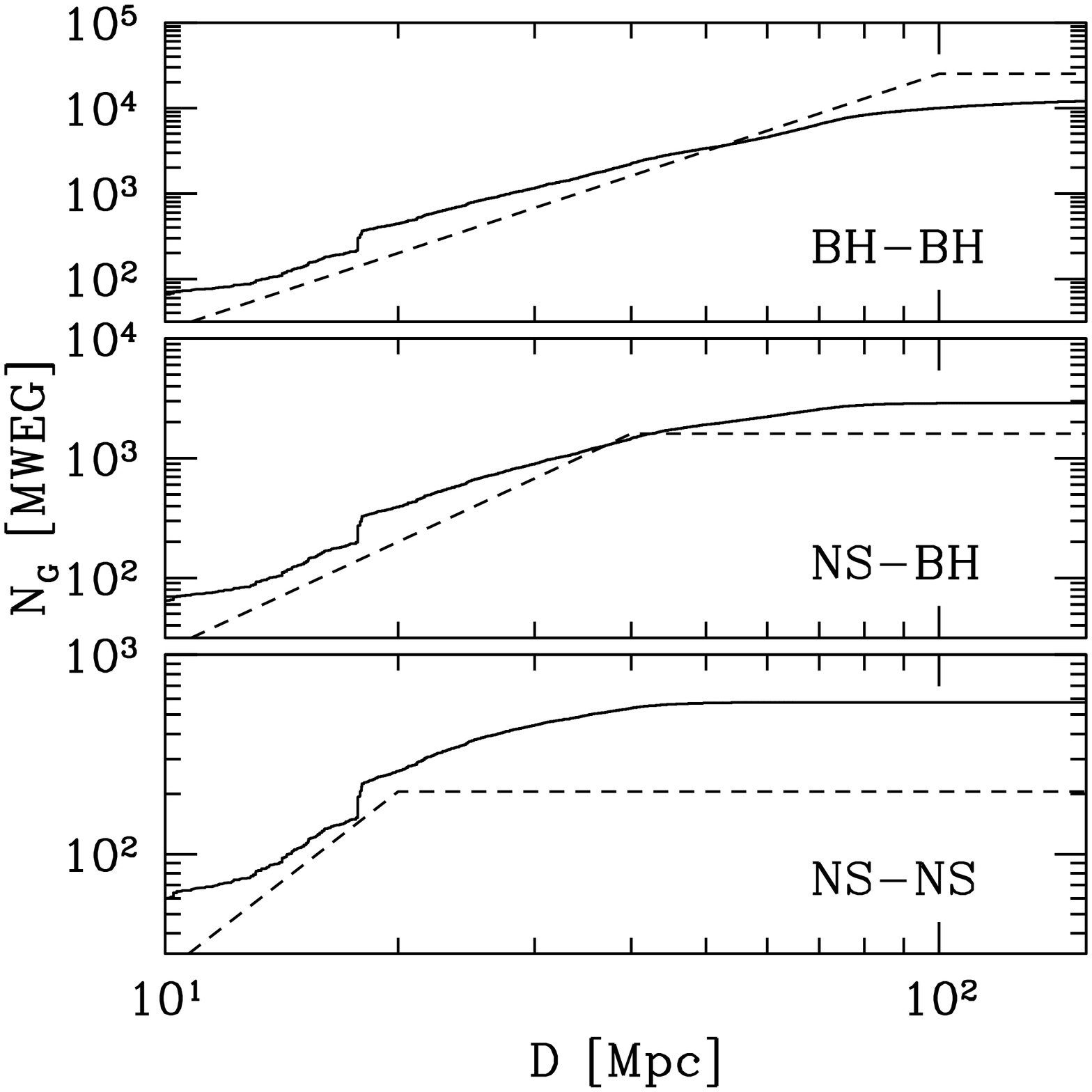,angle=0,width=6.2in}} 
\figcaption{The effective number of Milky Way Equivalent Galaxies
  (MWEG) surveyed by initial LIGO, for events within a distance $D$
  for BH-BH (top), NS-BH (middle), and NS-NS (bottom) binaries.  The
  dashed curves are calculated using the {\it uniform} approach
  adopting maximum detection distances for NS-NS, NS-BH, and BH-BH at
  20, 40, and 100 Mpc respectively.  The solid curves are calculated
  using the {\it galaxy catalog} approach.} 
\label{fig:SYNTHESIS}


\centerline{\psfig{figure=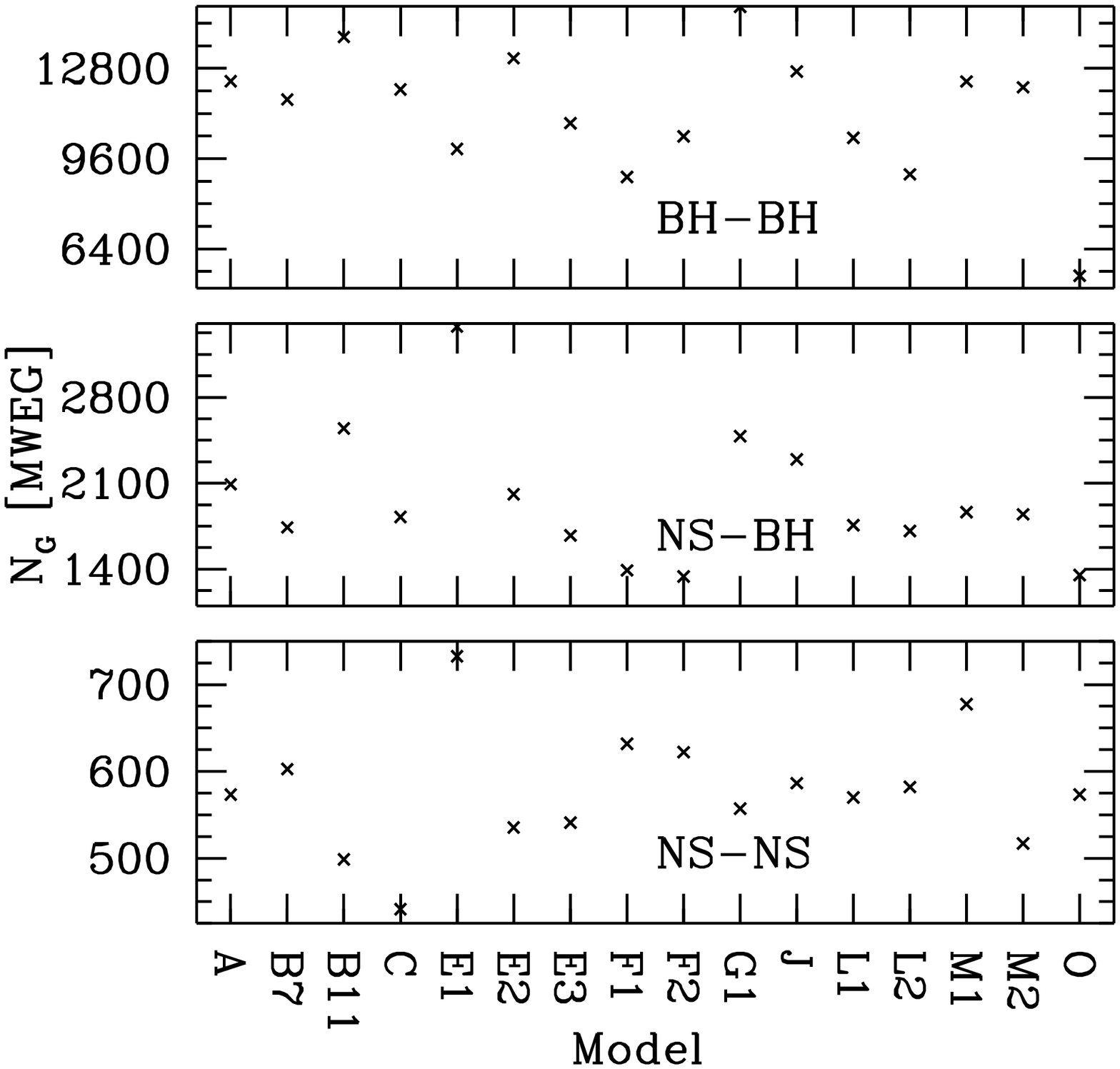,angle=0,width=5.2in}}
\figcaption{$N_{G}$ for Different Population Synthesis Models.
\label{fig:diff_models}}


 \centerline{\psfig{figure=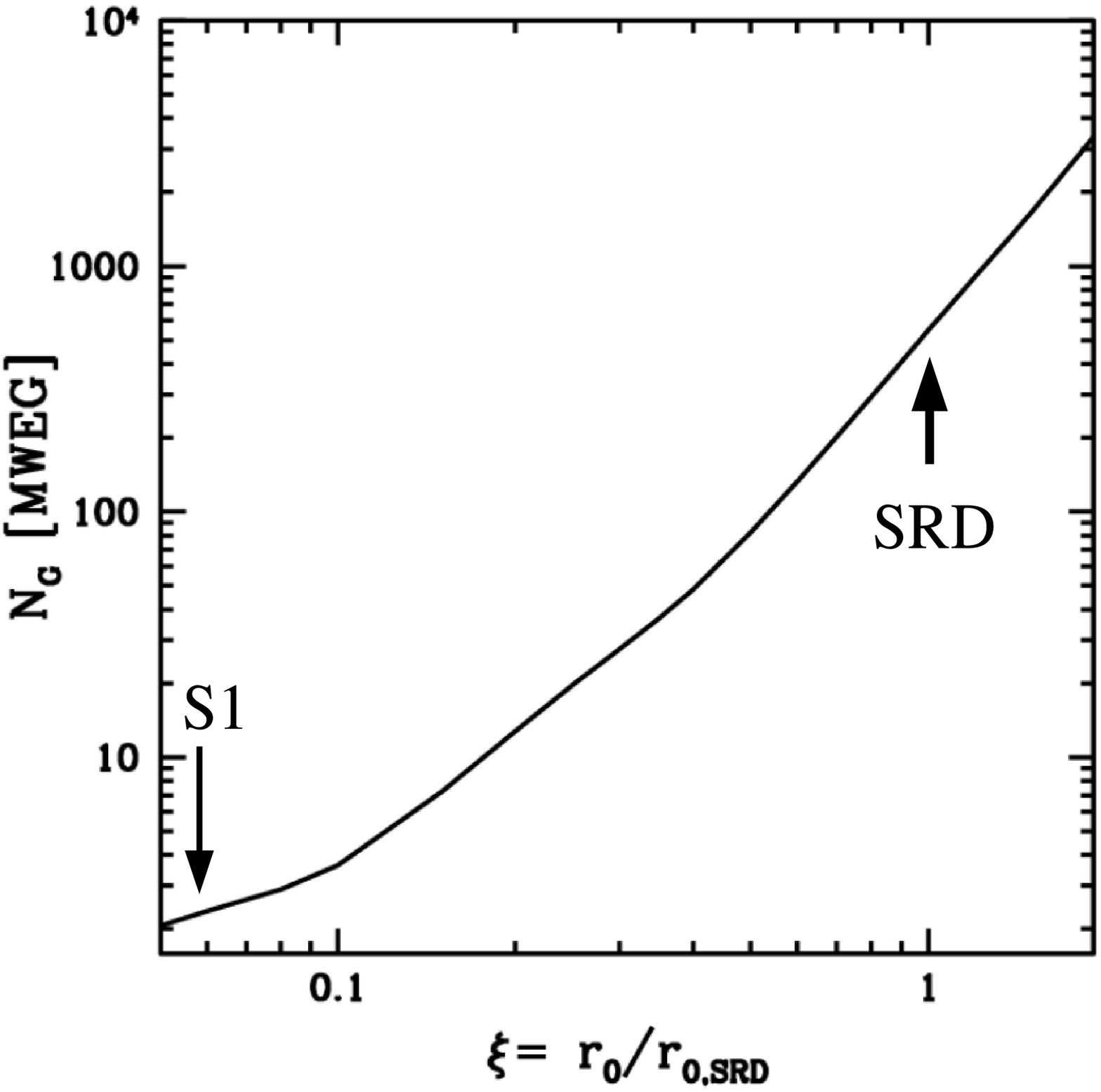,angle=0,width=5.2in}} 
\figcaption{The effective number of Milky Way equivalent galaxies
  surveyed by LIGO as a function of detector sensitivity $\xi$
  normalized to the initial LIGO design sensitivity for NS-NS
  binaries. The arrow on the left denotes the instrument sensitivity
  achieved during the first Science Run.} 
\label{fig:sensitivity}

\end{document}